\documentclass[preprint,showpacs,amsmath,amssymb,prb,aps]{revtex4}

\usepackage{graphicx}
\usepackage{dcolumn}
\usepackage{bm}

\begin{document}


\title{Mini-conference on angular momentum transport in laboratory and nature}

\author{Hantao Ji$^1$, Philipp Kronberg$^2$, Stewart~C. Prager$^3$, and Dmitri~A. Uzdensky$^4$}
\affiliation{Center for Magnetic Self-organization in Astrophysical and Laboratory Plasmas\\
$^1$Princeton Plasma Physics Laboratory, Princeton University, Princeton, New Jersey 08543 \\
$^2$Los Alamos National Laboratory, Los Alamos, New Mexico 87545\\
$^3$University of Wisconsin - Madison, Madison, Wisconsin 53706\\
$^4$Department of Astrophysical Sciences, Princeton University, Princeton, New Jersey 08540}

\date{\today}

\begin{abstract}

This paper provides a concise summary of the current status of the research and future perspectives discussed in the Mini-Conference on Angular Momentum Transport in Laboratory and Nature. This Mini-conference, sponsored by the Topical Group on Plasma Astrophysics, was held as part of the American Physical Society's Division of Plasma Physics 2007 Annual Meeting (November 12--16, 2007). This Mini-conference covers a wide range of phenomena happening in fluids and plasmas, either in laboratory or in nature. The purpose of this paper is not to comprehensively review these phenomena, but to provide a starting point for interested readers to refer to related research in areas other than their own.

\end{abstract}

\maketitle

\section{Introduction}

Understanding mechanisms of rapid angular momentum transport has been an important plasma physics topic for both laboratory and astrophysical plasmas. In high-temperature fusion plasmas, spontaneous spin-ups without apparent momentum inputs have been observed experimentally in many devices. Rotation and plasma flow in general are closely tied to transport of heat and particles across field lines. In astrophysical plasmas, angular momentum transport is central to accretion disk physics, such as star formation processes, mass transfer between binary stars, and active galactic nuclei. Angular momentum transport is also important to understand other astrophysical phenomena, such as the solar dynamo. Back on the earth, angular momentum transport can also be important in phenomena such as formation and maintenance of atmospheric cyclones (hurricanes and tornados), faster spin of the inner core than the outer core, and the related geodynamo.
 
There has been significant progress in the past decades in all of these areas in our understanding of these important phenomena, based on observations, theory, simulation and laboratory experiments. The Mini-conference on Angular Momentum Transport in Laboratory and Nature, held as part of the American Physical Society's Division of Plasma Physics 2007 Annual Meeting,
was intended to provide a platform to summarize and discuss the recent progress and future perspectives in this field from communities of plasma physics, astrophysics and geophysics.
Because of enthusiastic responses to the call, the number of sessions had to be increased from
two to four. Below we summarize each session primarily by its chair person in the following sequence:
angular momentum transport in neutral and electrically conducting fluids (Session I, chaired by H. Ji), astrophysical and laboratory plasmas (Session II, chaired by D. Uzdensky), fusion plasmas (Session III, chaired by S. Prager), and earth cores, sun, and laboratory (Session IV, chaired by P. Kronberg).
We note that this paper intends to provide only brief summaries of the contributions in these sessions without attempting to be a comprehensive review of any sort.

\section{Session I: Angular momentum transport in neutral and electrically conducting fluids}

This session was devoted to angular momentum transport phenomena in both 
electrically insulating and conducting fluids both in nature and laboratory. 
It began with a presentation by D. Nolan from the University of Miami
on angular momentum circulation in tropical cyclones such as hurricanes.
These strong atmospheric vortices are driven by the potential energy due to sufficiently large temperature differences between upper atmosphere and sea surface, as in a \lq\lq heat engine\rq\rq~\cite{emanuel03}. Hurricanes are formed due to spin-up of rapid vertical flows due to conservation
of absolute angular momentum (self angular momentum around their axes plus rotation of the
atmosphere). Transport of angular momentum within a hurricane is dominated by
global circulations through swirling boundary layers or Ekman layers, where the friction with
sea or land surface is a main loss of angular momentum~\cite{nolan05}. A typical hurricane
outside its core is anti-cyclonic, i.e., vorticity has an opposite sign to the rotation, and more specifically,
angular velocity is a decreasing function of radius while specific angular momentum increases
with radius. These flows belong to a rotating flow class, dubbed as \lq\lq quasi-keplerian" flows, since  they include keplerian flows typical of astrophysical accretion disks (see more below).
Interestingly, intense hurricanes have a flow profile often close to the keplerian flows, but probably
due to different reasons originated in their boundaries. Some internal waves can transport angular momentum within the hurricane vortex. The vortex-Rossby waves are considered to generate the observed spiral bands~\cite{montgomery97}, 
which in turn transport angular momentum radially inwards up an angular
velocity gradient, as also argued in accretion disks~\cite{balbus03}. In contrast, gravity waves
can transport angular momentum radially outwards but only very inefficiently.
The absence of efficient mechanisms to disperse the concentrated angular momentum explains the prolonged lifetimes of tropical cyclones.

P. Marcus from the University of California, Berkeley, presented a recent theory on
transporting angular momentum by 3D coherent vortices in protoplanetary disks~\cite{barranco05,barranco06}, rather than conventionally assumed turbulence~\cite{balbus98}. 
Vertically stratified disks support stable internal gravity waves, which can be generated by
noise near the disk mid-plane. These gravity waves propagate off mid-plane, then they
break to form long-lived 3D vortices. Constraints on certain symmetries of these coherent vortices can
lead to radial transport of angular momentum at a sizable rate. In addition to transporting angular
momentum, these coherent vortices can provide homes for planetesimal formation out of dust grains.
Various forces within the vortices can affect dust dynamics, but large pressure forces and vertical
circulations can help trap dust.
This is in contrast to turbulence, which is another candidate to transport angular momentum. But
it is difficult for dust grains to settle or aggregate into planetesimals in either hydrodynamic
or magnetohydrodynamic turbulence.

E. Kim from the University of Sheffield, UK, presented a summary of a series of recent theories
on turbulent transport of momentum, taking into account different combinations of effects
of rotation, stratification, flow shear, and magnetic field. Quasi-linear analyses were
used to calculate these effects, which are dynamically coupled through the equation
of motion and the induction equation. A common feature of these
effects is that they introduce anisotropy into the underlying turbulence, which in turn gives
anisotropic turbulent transport. Flow shear was shown to be able to strongly modify
turbulence, leading to weakened, anisotropic transport of momentum~\cite{kim05,kim06,kim07}.
In addition to the flow shear, when the fluid is strongly stratified and magnetized~\cite{kim07b},
or when fast rotation is introduced~\cite{kim07c}, momentum transport can become anti-diffusive, or
transported up the flow gradient. These offer possible mechanisms to form and sustain
sheared flows, which may have some similarities with the observed solar tachocline.

E. Schartman from Princeton Plasma Physics Laboratory presented recent results from
the Magnetorotational Instability (MRI) experiment at Princeton. There have been two
main proposed mechanisms to explain angular momentum transport accompanied with
fast accretion rate in astrophysical disks. When disks are hot, and thus sufficiently
ionized to conduct electric current, the MRI is destabilized to efficiently transport
angular momentum~\cite{balbus98}. In sufficiently cold disks, however, MRI can be
stable while hydrodynamic turbulence could arise due to nonlinear instabilities
at extremely large Reynolds numbers~\cite{richard99}. Investigations of nonlinear
hydrodynamic instability have been the first focus of the Princeton MRI experiment~\cite{ji01}.
By controlling boundary effects using two pairs of separately rotatable rings
at its axial ends~\cite{burin06}, no signs of hydrodynamic turbulence have been detected~\cite{ji06}
up to $Re=2\times 10^6$. Furthermore, by directly measuring the Reynolds
stress responsible for angular momentum transport using a synchronized 
dual laser doppler velocimetry system, the measured transport efficiencies are
well below the required values to be consistent with astrophysical observations.
Thus, the pure hydrodynamic turbulence has been effectively eliminated 
as a viable mechanism to transport angular momentum in accretion disks.
Initial results from the MHD experiments using liquid gallium have shown that 
a non-axisymmetric instability arises when a sufficiently strong, axial magnetic
field is applied to hydrodynamically turbulent flows.

F. Cattaneo from the University of Chicago presented results from state-of-art numerical 
simulations of MHD turbulence driven by MRI in a cylindrical geometry similar to the Princeton
MRI experiment. He pointed out that numerical experiments and laboratory experiments
are often complimentary to each other in the sense that laboratory experiments can often
achieve parameters beyond what can be achieved numerically, while numerical experiments
have great flexibilities in exploring wide ranges in parameter space. The new
results presented in this talk were obtained at magnetic Reynolds numbers much
higher than the experimentally achievable values, but at Reynolds numbers much 
below
the experimental values. It was shown that MHD turbulence develops due to MRI, and
the flows act as efficient dynamos because the turbulence seems to persist as long as the
flow is driven, even in the absence of an externally applied magnetic field. However,
the amplified magnetic fields have relatively small-scale structures, similar to what is
predicted for the small-scale dynamos, rather than the structures predicted for
the large-scale dynamos. 
Angular momentum is transported mostly by Maxwell stress, and, as a result, the background
flow profiles are strongly modified.

F. Stefani from Forschungszentrum Dresden-Rossendorf, Germany, presented
experimental results from the PROMISE (Potsdam ROssendorf Magnetic InStability 
Experiment) device in helical magnetic fields.
The experiments were motivated by the theoretical prediction that MRI can be 
destabilized at much smaller magnetic Reynolds numbers when a helical 
magnetic field (with both azimuthal and vertical components) is 
imposed~\cite{hollerbach05}.
The predicted traveling wave patterns, and their axial traveling direction, 
have been observed in the experiment 
by using ultrasonic diagnostics~\cite{stefani06,ruediger06,stefani07}.
Phase speeds of the traveling waves are also comparable, although not identical, 
to the predictions. Effects of radial boundary conditions have also been studied.
Axial boundary conditions have been further optimized in 
the modified experiment, PROMISE 2.
Initial results indicate that Ekman circulations are reduced, and
the measured phase speed of the traveling waves agrees better with the numerical 
predictions. 
He also briefly discussed whether this helical MRI can be applicable
for astrophysical disks, as well as the distinction between convective
and absolute instabilities when evaluating effects due to axial boundary 
conditions.

W. Liu from Los Alamos National Laboratory presented axisymmetric numerical and analytical 
studies~\cite{liu06b,liu07} of the helical MRI  in a cylindrical geometry 
similar to the PROMISE experiment. It was shown that helical MRI is a weakly 
destabilized hydrodynamic inertial oscillation propagating axially along the 
background Poynting flux in vertically infinite or periodic cylinders.
This mode is stable in Keplerian flow profiles regardless of axial boundary 
conditions. 
When the finite axial dimension is taken into account, the growth rates are 
reduced or the mode can be completely stabilized in inviscid analyses.
When the actual experimental boundary conditions are accurately modeled
in resistive and viscous simulations, 
the measured wave patterns and their amplitudes are successfully reproduced.
Contrary to previous claims, the waves are shown to be transiently amplified 
disturbances launched by viscous boundary layers, rather than globally 
unstable magnetorotational modes.

\section{Session II: Angular momentum transport in astrophysical and 
laboratory plasmas}

The second session of the mini-conference 
was mostly devoted to various astrophysical topics,
including related laboratory experimental work.

E.~Zweibel (University of Wisconsin - Madison) discussed the angular 
momentum transport problem in environments with very weak magnetic 
fields, relevant to various astrophysical contexts. 
She first addressed MRI in the weak field limit~\citep{krolik_zweibel06}, 
appropriate for seed magnetic field generated by the Biermann battery 
effect (with $\omega_{ci}/ \Omega \sim 1$).
She found that collisional disks are unfavorable for the weak-field MRI, 
because of strong resistive damping, but collisionless disks are more 
favorable, due to weaker damping and rapid growth on large scales because 
of the pressure anisotropy. 
She then discussed another topic of significant astrophysical interest 
where weak magnetic fields may have a strong effect on momentum transport: 
turbulent mixing and entrainment in magnetized shear flows, e.g., in jets. 
She reported the results of her ongoing numerical work with collaborators 
on studying the Kelvin-Helmholtz instability with a background magnetic 
field parallel to the flow in a resistive inviscid fluid~\cite{palotti08}. 
Contrasting 2D hydrodynamical and 2D MHD simulations, they found that the 
flow structure is much more intricate and the turbulent mixing layer spreads 
much stronger (2-3 times) in the MHD case. The flow becomes very chaotic, 
generating a rich magnetic structure and transiently amplifying the field 
(before it eventually decays due to resistivity). Momentum transport is also 
larger in the MHD case because of Maxwell stresses. However, tracer particle 
transport (chemical mixing) was similar in the two cases.

P.~Sharma (University of California, Berkeley) presented his numerical work 
on heating and angular momentum transport in hot accretion flows onto black 
holes (BHs) with low accretion rate, such as Milky Way's Galactic Center~BH. 
These systems (called ``radiatively inefficient accretion flows'' or RIAFs) 
have very low radiative efficiency, $L\ll \dot{M} c^2$, and very low density,
so understanding them requires analyzing the MRI in the collisionless regime.
To do this, one has to take into account the anisotropic pressure tensor, 
related to the anisotropic Braginskii viscosity. This requires formulating
a closure prescription for the heat fluxes that control the pressure 
anisotropy $\Delta P \equiv  P_\perp - P_\parallel$, e.g., as in kinetic MHD. 
This can be done once one incorporates small-scale instabilities that limit 
the growth of~$\Delta P$, such as the ion-cyclotron and mirror instabilities 
for ions and electron-whistler for electrons for $\Delta P>0$, and the 
firehose instability for $\Delta P<0$. The resulting prescriptions allow
one to perform a linear stability analysis for the collisionless MRI and 
do shearing box simulations~\citep{sharma06}. A large $\Delta P$ could, 
in principle, suppress~MRI, but it is itself suppressed by these small-scale 
instabilities to $|\Delta P|/P \sim 1/\beta \ll 1$. Then, MRI is not strongly 
affected and proceeds essentially as in the MHD regime. Sharma also reported 
a significant (a few per cent of the total energy) electron heating by 
the electron-whistler instability, which is important because it ultimately 
determines the radiative efficiency of the plasma~\citep{sharma07}. 

P.~Kronberg (Los Alamos National Laboratory and University of Toronto) discussed 
observational evidence for magnetized accretion in super-massive BHs 
in active galactic nuclei (AGN), based on studies of giant radio jets 
and lobes. Simple estimates suggest that a significant (20\%) fraction 
of the BH gravitational infall energy feeds into these structures, which 
magnetizes the inter-galactic medium (IGM) and has important effects on its 
evolution~\cite{kronberg01}. He then discussed observational evidence for 
ongoing cosmic ray (CR) acceleration in the lobes of Mpc-size radio sources. 
The evidence comes from the relatively short inferred synchrotron lifetime 
of CR electrons compared with their travel time across these structures. 
It suggests a gradual ($10^8-10^9$ years) dissipation of magnetic energy, 
presumably by reconnection. Kronberg suggested that substantial, although 
still undetected, magnetic energy may be present in the~IGM, and that it 
may be detected in the future as faint diffuse synchrotron radiation. 
He discussed difficulties in observational jet plasma diagnostics, and 
some ways to address them, e.g., using Faraday rotation and the total 
synchrotron luminosity measurements in knots. He presented the first 
preliminary measurement of an electric current in an extragalactic kpc-scale 
jet ($\sim 7 \times$10$^{17}$ amperes), and its sign -- directed 
{\it away} from the AGN~BH.   

E.~Liang (Rice University) discussed a topic related to Sharma's ---
MRI-induced turbulent heating in low-luminosity BH systems and the 
resulting non-thermal emission, including very hard $\gamma$-ray-spectrum. 
This ongoing collaborative project attempts to reach across the huge 
dynamic range of this complicated problem and to connect global 2D General Relativistic (GR) 
MHD simulations of MRI with studies of small-scale turbulent dissipation and 
finally with radiative transport. The overall physical picture is the
following: MRI produces fully developed MHD turbulence; the energy cascades 
down to small (unresolved) scales, leading to plasma heating and non-thermal 
particle acceleration. Non-thermal electrons produce soft (IR) synchrotron 
emission which is then Compton-scattered by the same non-thermal electrons 
to yield the hard $\gamma$-ray emission. Finally, non-thermal ions produce 
very hard $\gamma$-ray emission via pion production and decay. 
Liang and his collaborators plan to employ a set of theoretical 
and numerical tools. On largest scales, they use GRMHD simulations 
of MRI turbulence. Then, to study the dissipation of the turbulent 
cascade and particle energization at small scales, they plan to solve 
the Fokker-Planck equation and/or use PIC codes. The goal of this stage 
is to derive a sub-grid model for both thermal heating and non-thermal 
electron and ion spectra, to be used locally at each grid zone in the 
large-scale simulation. Once this is done, the computation will be coupled 
to a 2D Monte-Carlo radiation transfer code that includes all the important 
photon production and scattering processes~\citep{boettcher03}. As a result, 
one will be able to predict the spectra and even build images of the accretion 
flow for different viewing angles. 

B.~Coppi (MIT) presented his work on the ``thermo-rotational instability'',
a vertically localized MHD ballooning mode driven by vertical gradients of plasma density and temperature in stratified differentially rotating accretion disks~\cite{coppi08}.
Quasi-linearly, the main effect of this instability is to drive the vertical profiles of the plasma density and temperature towards the adiabatic condition $\eta_T \equiv (d\log{T}/dz)/(d\log{\rho}/dz)=2/3$, as has been shown previously~\cite{coppi_keyes03}. In addition, Coppi suggested that the nonlinear development of the thermo-rotational mode can result in a break-up of the accretion disk into a sequence of counter-streaming toroidal current channels
--- axisymmetric plasma rings~\cite{coppi_rousseau06}.

D.~Homan (Denison University) described the MOJAVE project, a systematic 
study of extra-galactic jets using VLBA. He addressed the question of what 
these radio observations can tell us about the helical magnetic field 
structure in AGN radio jets on parsec scales (e.g., Cyg.~A). Thus, gradients 
of Faraday rotation measured across the jet represent evidence for toroidal 
magnetic fields~\citep{attridge05}, while the observed sign-consistency of 
circular polarization (on timescales of up to 20 years) suggests a persistent 
ordered net magnetic flux along the jet axis~\cite{homan01}. 

S.~Prager (University of Wisconsin - Madison) talked about momentum 
transport by current-driven instabilities (in contrast to the MRI 
flow-driven transport). He first described the ongoing experimental 
work on Madison Symmetric Torus (MST), a reversed-field pinch
\citep{hansen00}.
The plasma is observed to rotate differentially in MST, with periods 
of steady differential rotation being disrupted by very rapid (compared 
to classical) momentum transfer events, showing a sowtooth-like evolution 
of the plasma core toroidal velocity. The rotation profile flattens and 
angular momentum is transported outward. Simultaneous magnetic measurements 
indicate that these events coincide with fast magnetic reconnection (both at 
the core and at the edge). The leading explanation is current-driven tearing 
instability that involves the reconnection of the reversing toroidal field.
This produces islands with both radial and toroidal field, and since these 
islands are immersed in a background shear flow, angular momentum is 
transported via Maxwell (and Reynolds) stresses. Even a single tearing 
mode transports toroidal momentum radially. But with multiple modes, 
the transport is amplified by nonlinear coupling. Prager presented a 
comparison of the properties of tearing and MRI mechanisms in relation 
to momentum transport. To understand the roles of tearing mode in momentum 
transport, the MST group employs a combination of theoretical and 
experimental work. On the theoretical side, a quasi-linear calculation 
was done~\cite{ebrahimi07}, as well as 3D resistive MHD simulations 
(with an artificial force driving the shear flow), including the studies 
of multi-mode nonlinear coupling. Prager also discussed astrophysical 
implications of current-driven instability for momentum transport, e.g., 
for thick disks in the strong-field case, which is more relevant for jets 
than for disks.

C.~Forest (University of Wisconsin - Madison) proposed a new type 
of unmagnetized plasma experiment aimed at studying MHD dynamo and the MRI. 
He pointed out that the existing liquid metal experiments are limited
to small magnetic Prandtl numbers due to material properties, 
and small magnetic Reynolds numbers due the limited stirring power. 
In addition, liquid metals are not 
transparent and so are difficult to diagnose. Plasmas experiments
can provide more flexibility in terms of accessible magnetic Prandtl number, 
do not require as much power, enable one to study non-MHD 
plasma effects, and are easier to diagnose. However, such laboratory plasma 
studies are actually difficult because strong magnetic fields are usually
required to insulate the hot plasma from the cold walls of the vacuum vessel. This makes 
traditional plasma devices unsuitable for studies of MHD dynamo and MRI. 
In his talk, Forest proposed an unconventional experimental set-up involving
a large number of axisymmetric ring cusps made of strong permanent magnets 
lining the inner wall of a spherical chamber. The magnetic field is then 
very strong only close to the wall, providing magnetic insulation while
leaving the bulk of the interior volume relatively unmagnetized. Forest 
then outlined possible ways to drive the plasma flow and presented estimates 
demonstrating the viability of this concept. He discussed several possible 
experimental studies that could be performed with this device, including 
spherical von-Karman flow, small-scale dynamo at large Prandtl number, 
and a study of the MRI. He also described a small pilot experiment to 
study a plasma Couette flow now under construction in Madison.

The last talk of the session was given by T.~Islam (\'Ecole Normale Sup\'erieure, France, and University 
of Virginia). He talked about the magneto-thermal and magneto-viscous 
instabilities (MTI and~MVI, respectively) and their role in hot, dilute, 
nearly collisionless plasmas in low-luminosity accretion flows on 
super-massive black holes in the centers of most inactive galaxies, 
including the Milky Way (Sag.~A*). 
He first explained the basic physical mechanisms of these instabilities 
and the key role played by even a weak magnetic field in making both 
the heat conduction and the viscous stress tensor highly anisotropic 
\citep{balbus01,balbus04,islam_balbus05}. He then compared the linear 
dispersion relations and the quadratic quasi-linear heat and momentum
fluxes for collisional MTI and collisionless~MTI, and underscored the 
need for improved Landau fluid closures, similar to the collisionless 
MHD closures\citep{snyder97}. He also discussed the existing numerical 
studies of the effect of MTI turbulence on angular momentum transport 
in accretion flows.

\section{Session III: Angular momentum transport in fusion plasmas}

Momentum transport and momentum generation are important physics issues for tokamak plasmas.  Momentum transport is observed to be more rapid than the classical value. In addition, tokamak plasmas rotate even in the absence of an applied torque.  The cause of the transport and the spontaneous, or intrinsic, rotation are yet undetermined.  The presence and structure of plasma rotation is of substantial consequence.  Sheared flow can reduce turbulence, influence the transition to the H mode, form internal transport barriers, and stabilize resistive wall instabilities.  A session during the miniconference was dedicated to laboratory observations of momentum transport and intrinsic rotation, as well as theoretical studies.

J. Rice (MIT) inferred general properties of rotation in tokamaks by comparing observations from many tokamaks in plasmas in the high confinement (H) mode~\cite{rice07}.  The intrinsic rotation appears to propagate inward from the edge, with rotation being larger in H mode plasmas (relative to L mode).  Rotation increases with stored plasma energy (normalized to the plasma current).  Both the thermal and Alfv\'en Mach numbers increase with the normalized beta, possibly implying a connection to MHD.  In ITER (International Thermonuclear Experimental Reactor), the normalized beta is expected to be about 2.6, so that the thermal and Alfv\'{e}n Mach numbers would be expected to be about 0.3 and 0.02, respectively, with rotation speeds about 250 km/s.  This value of rotation might be large enough for stabilization of resistive wall instabilities.  Rotation is roughly independent of the normalized gyroradius and normalized collisionality, possibly suggesting independence from electrostatic turbulence.  

Experimental studies in the DIII-D tokamak have also revealed how momentum transport (or the momentum confinement time) varies with key plasma parameters. W. Solomon (PPPL) presented such information for DIII-D plasmas with only a weak torque provided by neutral beams~\cite{solomon07}.  The torque in H mode plasmas with the presence of edge-localized-mode (ELM) was scanned, holding beta constant.  The intrinsic rotation profile is deduced from the scan.  The intrinsic rotation, of interest for its own sake, must also be taken into account in evaluation of the momentum confinement time.  It is found that the confinement time decreases as the applied torque increases.  A cautionary note to the above conclusions is that they can be influenced by anomalous diffusion of fast ions (possibly implied by neutron emission measurements). 

In the NSTX (National Spherical Torus Experiment), a small aspect ratio tokamak, strong shear flows are observed~\cite{kaye07,solomon07b} that are $5 Ð 10$ times larger than the growth rate for ion temperature gradient instabilities, as reported in a paper by S. Kaye et al (and delivered at the miniconference by W. Solomon).  The momentum diffusivity behaves quite differently than the ion particle diffusivity.  The momentum diffusivity is larger than the neoclassical value in plasmas for which the ion particle diffusivity is approximately neoclassical.  Moreover, the momentum diffusivity scales differently than particle diffusivity (unlike the case of normal aspect ratio tokamaks). For example, the momentum diffusivity scales more strongly with the toroidal magnetic field than with the plasma current, unlike the particle diffusivity.  Momentum perturbation experiments were performed by imposing rotation damping by magnetic perturbations (with toroidal mode number = 3).  The rotation damping is consistent with neoclassical toroidal viscosity.  Transport analysis indicates the presence of an inward momentum pinch.

G. Tynan from UCSD presented experimental results from in a linear plasma device 
on the development of an azimuthally symmetric, radially sheared plasma flow in the absence of an external source of momentum.  The measured radial flux of azimuthal momentum (i.e., the Reynolds stress) is consistent with such a sheared flow, which is damped by ion-neutral drag and ion-ion collisional viscosity~\cite{tynan06,holland06}. This indicates that turbulent momentum transport appears to be the mechanism that sustains the shear flow.  Two-field turbulence simulations of the experiment also show the natural emergence of such a shear layer from nonlinearly interacting collisional drift waves, providing further support to this interpretation.  It is also found that the shearing rate  varies on a very slow frequency scale (i.e. $\sim$ 1\% of the diamagnetic drift frequency), and is linked to a similar slow variation in the turbulent Reynolds stress and the turbulence amplitude.  In addition, radially outgoing bursts of plasma are occasionally born in the region of the shear layer, and then propagate convectively outwards, away from the main plasma~\cite{antar07}.  These recent observations suggest that shear layer modulation, turbulence and turbulent Reynolds stress modulation, and generation of bursts of outward going plasma events (referred to as blobs, fingers, and avaloids in the literature) are linked, and form a complex dynamical system.

Experimental studies of rotation have also been reported in a simple torus, the TORPEX experiment. TORPEX contains a dominant toroidal magnetic field and a much weaker ($\sim$ 100 times) vertical field.  Thus, the field lines are open.  The plasma is produced by microwaves, and the resulting plasma exhibits both vertical and toroidal flows.  The 2D flow pattern (reported by B. Labit) has been measured with an insertable Mach probe.  The flow pattern agrees with that expected from $E \times B$ drift in the measured radial electric field. However, the torodial flow is about 100 times larger than expected from the $E \times B$ drift (and comparable in magnitude to the vertical flow).  Local increases in toroidal flow are observed simultaneous with density blobs.  These features are thought to be associated with interchange modes~\cite{furno08}. The excess flow and the density blob are both expelled radially and simultaneously, suggesting a conservation of momentum. 

Substantial theoretical effort is underway to explain both the intrinsic rotation and the transport of momentum. Three theoretical studies were presented at the mini-conference, spanning effects from turbulent transport to neoclassical momentum generation.  {\"O}.D. G{\"u}rcan from UCSD (delivering a paper by P. H. Diamond et al.~\cite{diamond07}) presented a theory of turbulent equipartition of angular momentum~\cite{gurcan07,hahm07}.  The theory is motivated, in part, by experimental observations that the rotation profile is centrally peaked in some cases, and that the Reynolds stress and rotation are correlated (implying an anomalous mechanism). From conservation of angular momentum density, a model is derived in which turbulent mixing leads to conservation of a local quantity $L_\Phi/\lambda$, where $L_\Phi$ is the toroidal angular momentum, and $\lambda=B$ for a slab with inhomogenous magnetic field, or 
$\lambda=B^2$ for a low $\beta$ torus.  The mixing stops at Òequipartition,Ó which occurs when the angular momenta in two adjacent toroidal shells are equal.  This process leads to an inward pinch of angular momentum density.   A general model including both the effects of $E\times B$ shear driven residual stress~\cite{gurcan07b} and the turbulent equipartition pinch is introduced. Intrinsic rotation and centrally peaked profiles are possible consequences of this model.

A. Aydemir presented a neoclassical theory for edge flows and intrinsic momentum~\cite{aydemir07a,aydemir07b}. 
The model is motivated by three experimental observations: all tokamaks have mass flow at the plasma edge, all tokamaks have toroidal rotation in the core, and the L-H transition requires more power if the $\nabla B$ drift points away from the x-point in the magnetic field.  The neoclassical model might be able to explain all three observations.  In the presence of collisions, neoclassical theory predicts that a dipolar flow pattern develops in which a major radial flow arises directed from the low field side to the high field side, leading to a return flow in the scrape-off layer. This constitutes a dipole pattern of counter-rotating vortices, or transport-driven flows.   The interaction of these flows with a poloidally asymmetric field geometry (e.g., due to a divertor null in the poloidal field) can provide a net toroidal momentum input from the edge, a possible source of intrinsic rotation.  The flows have the correct symmetry properties and dependence on field direction to account for the $\nabla B$ drift dependence of the L-H power threshold.

C-S Chang described the sources for angular momentum in the tokamak edge plasma.  Full-f gyrokinetic ion-electron simulation~\cite{ku06} of quiescent edge plasmas, (including the pedestal, lower single null magnetric separatrix, scrape-off and the grounded material wall,) finds three spontaneous rotation sources in the edge plasmas.  A) Ion orbit losses due to the existence of a magnetic x-point~\cite{chang04,chang02} gives rise to co-current rotation in the H-mode like pedestal top/shoulder region. In an L-mode like pedestal, the co-current rotation source becomes small~\cite{chang06}.  B) A strong negative radial electric field in the vicinity of the magnetic separatrix yields a sharp valley of counter current rotation at the same radial location.  In an H-mode pedestal, the negative radial electric field  and, thus, the counter-current rotation valley, becomes strong.  In an L-mode pedestal, this phenomenon becomes weaker.   C) Plasma wall interaction generates positive radial electric field and co-current plasma rotation in the scrape-off region. The co-current rotation in the scrape-off plasma becomes stronger in an L-mode edge.  The general trend is that in the lower single null L-mode plasma, there is a strong co-current rotation in the scrape-off region, without much rotation in the pedestal top.  As the pedestal steepens, the co-rurrent rotation moves into the pedestal top from scrape-off plasma. These findings from the gyrokinetic particle simulation are consistent with experimental observations~\cite{rice04}. It has also been shown that plasma interaction with background neutral particles can modify the toroidal rotation~\cite{maingi04}.  When the neutral gas is puffed from the high magnetic field side in NSTX, the rotation becomes stronger and the H-mode power threshold is lower than when the gas puff is from the low magnetic field side.  Resonant magnetic perturbations are also seen to produce spontaneous co-current rotation. 

\section{Session IV: Angular momentum transport in the Earth's core, Sun, and laboratory}

Session IV began with a presentation by B. Buffett (University of Chicago) on angular momentum transport and magnetic interactions between the Earth's cores and the mantle. 
Magnetic field is continuously generated by convection in the liquid iron core, known as the geodynamo~\cite{buffett00,roberts00},
and is believed to sustain a differential rotation between the inner and outer boundaries.
The differential rotation is thought to be a consequence of large-scale fluid circulation in a region defined by the tangent cylinder (a hypothetical cylinder that is tangent to the solid inner core at the equator). Different detailed mechanisms were discussed on the interactions between gravitational and magnetic forces that regulate the deformation, rotation, and axial misalignment of the Earth's core. 

D. Lathrop (University of Maryland) summarized several experiments that study flows between differentially rotating spheres and cylinders, in attempts to model geophysically relevant phenomena. 
In a spherical Couette flow of liquid sodium, MHD modes 
were observed when an axial magnetic field is imposed along the rotation axis~\cite{sisan04}. 
He attributed the modes to the MRI because of associated torque increases and similarities
in dependences on the field strength, although there were differences on the predicted
mode structures and the boundary conditions, as well as the fact that the base states were
already hydrodynamically turbulent before magnetic field is imposed.
He went on to describe experimental results on inertial waves driven by differential rotation 
in another (larger) spherical Couette flow also in liquid sodium~\cite{kelley07}. 
The magnetic field was used as a tracer measurement for the fluid motion.
Despite a turbulent background, coherent 
oscillations were observed which agreed with linear 
theory of the inertial waves, in terms of frequency, wave number,
and induced magnetic structure. He hypothesized that these waves are excited by
the differential rotation of the inner sphere to the outer sphere, and similar phenomena
may exist in the Earth as the inner core rotates faster relative to the outer core.

The tachocline layer of the solar interior is a puzzling phenomenon.
It has been a focus of several efforts to model and understand it. 
Its importance as a plasma physics testbed is underlined by the existence of a recent book 
on the subject ~\cite{hughes07}.
The tachocline is a thin 
($\leq $ 4 $\%$ of R$_{\odot}$) layer at r $\sim $ 0.7 R$_{\odot}$. It is very stably stratified.
It separates the inner, solid body rotating 
radiative zone from the outer convection zone, and appears to play a major role in transporting angular momentum inward to the sun's core. Most solar dynamo models rely on processes at or near the tachocline. The differential rotation which occurs throughout the convection zone occurs above, but not below this remarkable, thin discontinuity. It also influences the sunspot cycle behavior. An overview talk by S. Tobias (Leeds University, UK) presented various ideas on how angular momentum could be transported through the tachocline, and summarized different competing physical models~\cite{tobias07}. The tachocline has a relatively high density, 
$\rho \sim $ 0.2 g cm$^{-3}$ and is hot (T $\sim 2\times 10^{6}$K), giving
a pressure of $\sim$6 $\times$ 10$^{12}$ Pa. Its plasma $\beta$ is $\sim$ 10$^{5}$ with magnetic field in the range 10$^{4} - 10^{5}$G.  A definitive physical understanding has been elusive up to now, but it seems likely that mechanisms of turbulence in this thin layer hold the key to the tachocline's fundamentally important influence on the properties of the sun, and by extension the solar wind and sun-planet interactions in the inner solar system. A complete physical description the solar tachocline might simultaneously clarify how the turbulent dynamo in magnetized shearing flows could cause inverse cascades in other astrophysical situations where turbulence is dominant.

A talk along similar lines was given by Tamara Rogers (High Altitude Observatory and 
National Center for Atmospheric Research), who compared purely hydrodynamic with magnetized models to explain the relatively uniform rotation of the solar interior. She described a model in which the internal field, which need not be more than a few mG, is confined below the tachocline by the dynamics of the convection zone. In this model, by Gough and McIntyre~\cite{gough98}, the differential rotation in the convection zone is the 
driver of meridional circulation, which acts to confine the magnetic field that is embedded below.

J. Anderson and E. Kim (University of Sheffield, UK) examined the role that coherent structures play in momentum transport in a more general plasma context. They calculated probability distribution functions (PDF) for momentum transport, and find that the momentum distribution is better characterized by an exponential with wide outer tails, rather than a gaussian~\cite{kimej03}. They discuss some implications for shearing flows in astrophysical situations, by calculating the PDF's of the momentum flux and applying these to the formation of plasma shear flows.          

Related to the above, a comparably difficult puzzle involves asking how effectively the accretion disk corona, with its magnetized, low-$\beta$ hot plasma, can transfer angular momentum across the disk. D.~Uzdensky and J.~Goodman (Princeton University) described their new model which constructs the coronal field as a statistical ensemble of magnetic loops. The loops' wandering footpoints are at the turbulent disk and the corona evolves by a combination of (1) the randomly moving footpoints combined with (2) Keplerian shear, and (3) reconnection due to binary collisions of the coronal loops. They have been able to self-consistently compute the distribution of loop sizes and orientations, the height dependence of $<$B$^{2}$$>$/8$\pi$, the equilibrium shapes of loops, the energy associated with each loop, and the overall magnetic torque~\cite{uzdensky07,uzdensky08}. 
This model, which incorporates reconnection in loop-loop collisions, might also 
provide a future ``template'' for solar corona.     

Transfer of angular momentum is integral to the function of an accretion disk - jet system. It is the reason why gravitational collapse of stars, black holes, etc. can form. D.J. Ampleford (Sandia National Laboratories) described a laboratory experiment using a twisted conical wire array to produce a z-pinch~\cite{amplef07}. The Lorenz force accelerates plasma toward the cone axis. This twisted wire array configuration produces a radial current and an axial magnetic field component.  Both angular and axial momentum are conserved, and the former produces a conical shock that conserves angular momentum as the jet propagates away. In this experiment the ratio of azimuthal to axial velocity is measured, and is of order 10\%, similar to what has been measured in stellar jets~\cite{bacchiotti02}. The jet has a hollow density profile, and the conical shock radius increases as radative cooling proceeds in the course of the outflow. 

An experimental measurement of ion viscosity has not previously been done although predictions of its value, using
Braginskii's theoretical formulation~\cite{braginsk65}, have long been known. L. Dorf presented the first experimental evaluation of ion viscosity on behalf of a team that used the Reconnection Scaling Experiment (RSX) at LANL. They succeeded in measuring the axial flow velocity, density, electron temperature, and magnetic field components at two different axial locations~\cite{intrator07}. They verified a plasma flow velocity of order the ion acoustic speed, and that it decreased due to radial shear of the axial velocity, caused by ion viscosity. They were able to measure the ion temperatures at several radial locations, via spectroscopic measurement of the Doppler broadening. This new result is important for the design of magnetoplasmadynamic thrusters, and also for clarifying the physics of astrophysical jets.         

\section{Conclusion}

Angular momentum transport in fluids and plasmas is fundamental to the workings of the Universe, and hence our existence. Its understanding is also key to the future exploitation of new technologies in space and energy production (especially fusion science). This Mini-conference
has covered a wide range of phenomena related to this topic, and many of them
are traditionally separated in different fields.
We hope this concise summary can be used as a stepping stone for researchers in these
separate fields to find relevant references in fields other than their own.

A few observations can be made. First, many phenomena indeed share
commonalities even though they are different in spatial and temporal scales,
and are driven by different energy sources. The key here is dimensionless
parameters. By carefully matching these dimensionless parameters,
with appropriate initial and boundary conditions, the same physical processes
take place in different systems, whether they are natural phenomena,
laboratory experiments, or numerical experiments. This scalability of
physical phenomena is fundamental to the interplay between different
systems.

Second, there also exist large gaps in some of the dimensionless
parameters between different systems, and these gaps will not likely
disappear in the foreseeable future. This is most severe between
laboratory/numerical systems and natural phenomena, and less severe
but still problematic between laboratory experiments and numerical experiments.
Here, fourth and the most insightful system --- analytic theory ---
can play a key role to close these gaps to bring the different systems
to unified understandings. Although powerful, these theories do
not appear overnight. Often they are born from many failed attempts,
and more importantly from the interplay between different systems.
For example, physical intuitions of a large astrophysical system can be built 
based on laboratory and numerical experiments despite large disparities
in dimensionless parameters. These physical intuitions can accumulate into
a ground breaking theory, which can unify our understandings in
different systems.

We hope that this Mini-conference serves as a modest first step
for the interplay between different systems on the problem of angular
momentum transport.




\end{document}